\documentclass{ws-ijmpa}

\begin{document}

\markboth{T. LEITNER, U. MOSEL, L. ALVAREZ-RUSO} {IN-MEDIUM EFFECTS IN PHOTO- AND NEUTRINO-INDUCED REACTIONS ON
NUCLEI}

%
\catchline{}{}{}{}{}
%

\title{IN-MEDIUM EFFECTS IN PHOTO- AND NEUTRINO-INDUCED REACTIONS ON NUCLEI}

\author{T. LEITNER, U. MOSEL\footnote{electronic address: mosel@theo.physik.uni-giessen.de}, L. ALVAREZ-RUSO}

\address{Institut fuer Theoretische Physik, Universitaet Giessen, Germany}

\maketitle

\begin{history}
\received{Day Month Year}
\revised{Day Month Year}
\end{history}

\begin{abstract}
In this talk various aspects of in-medium behavior of hadrons are
discussed with an emphasis on observable effects. It is
stressed that final state interactions can have a major effect
on observables and thus have to be considered as part of the theory.
This is demonstrated with examples from photo-nucleus and neutrino-nucleus
interactions.

\keywords{hadrons in medium; photoproduction; neutrino scattering.}
\end{abstract}

\ccode{PACS numbers: 25.20.Lj, 25.30.Pt}

\section{Introduction}
Hadrons, embedded inside nuclei, obviously change some of their
properties. They acquire complex selfenergies with the real parts
reflecting the binding (or non-binding) properties and the imaginary
parts reflecting the interactions and possibly their changes inside
the medium. Particles that are produced through resonances or -- at
high energies -- through strings become physical, on-shell particles
only after some formation time. In this case the nuclear medium may
affect the formation process and can thus act as a  micro-detector
for the early stages of particle production.

Naively, one expects that in lowest order all in-medium effects go
linearly with the density of nuclear matter, $\rho$, around the
hadron. This has triggered a series of experiments with relativistic
and ultrarelativistic heavy-ions, which can reach high densities,
that have looked for such effects and have indeed reported in-medium
changes of the $\rho$ meson\cite{Ceres,Xu,NA60}. However, it has
been pointed out quite early\cite{Mosel} that also experiments with
microscopic probes on nuclei can yield in-medium signals that are as
large as those obtained in heavy-ion collisions. Although, of
course, the density probed here is always below $\rho_0$ the
observed signal is cleaner in the sense that it does not contain an
implicit integration over very different phases of the reaction and
the nuclear environment. The signal to be expected is also nearly as
large as that seen in heavy-ion collisions. This idea has been
followed up in recent experiments with photons on nuclei\cite{g7,Trnka},
where indeed changes of the $\omega$ meson in
medium have been reported\cite{Trnka}.

In Ref.~\refcite{Mosel1,Mosel2} we have discussed the relevant
questions and theoretical studies of in-medium properties in some
detail. Such calculations necessarily rely on a number of
simplifying assumptions, foremost being that of an infinite medium
at rest in which the hadron under study is embedded. In actual
experiments these hadrons are observed through their decay products
and these have to travel through the surrounding nuclear matter to
the detectors. Except for the case of electromagnetic signals
(photons, dileptons) this is connected with often sizeable final
state interactions (FSI) that have to be treated as realistic as
possible. For a long period the Glauber approximation which allows
only for absorptive processes along a straight-line path has been
the method of choice in theories of photonuclear reactions on
nuclei. This may be sufficient if one is only interested in total
yields. However, it is clearly insufficient when one aims at, for
example, reconstructing the spectral function of a hadron inside
matter through its decay products. Rescattering and sidefeeding
through coupled channel effects can affect the final result so that
a realistic description of such effects is absolutely mandatory.

In this talk we will give an overview of this field with an emphasis
on observable effects in photonuclear and
neutrino-induced reactions. More details can be found in two
previous reviews\cite{Mosel1,Mosel2}.

\section{In-medium effects}

The model we are using for the description of photon- und
neutrino-induced reactions factorizes into three ingredients. First,
there is shadowing in the entrance channel that comes about by a
quantum mechanical coherence effect. This is essential for photon
energies of about 1 GeV on upwards and for small virtualities
$Q^2$~\cite{Falter}. Second there is an elementary interaction of
the incoming probe with individual nucleons, the assumption being
here that the processes under study are all one-body processes. At
this stage also 'trivial' many-body effects, such as Fermi motion
and Pauli-blocking, can be taken into account.

\begin{figure}[tb]
\centerline{\includegraphics[width=7cm]{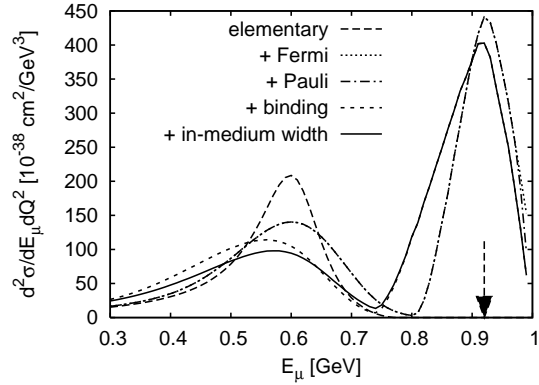}}
\caption{Inclusive double differential cross section $d \sigma/
dQ^2\,dE_{\mu}$ for charged current scattering of $\nu_{\mu}$ on
$^{56}$Fe at $E_{\nu} = 1$ GeV and $Q^2=0.15$ GeV$^2$ (from
Ref.~\protect\refcite{Leitner-CC}).} \label{total}
\end{figure}

How these effects influence the inclusive cross section is shown in
Fig.~\ref{total} for the example of neutrino scattering off a Fe
nucleus.
The left peak is due to $\Delta$ excitation, the right one to
quasielastic scattering. The dashed line shows the elementary cross
section for $\Delta$ production. The position of the
$\delta$-function of the QE cross section is indicated by the arrow.
We subsequently include Fermi motion and Pauli blocking as well as
the binding of the nucleons in a mean-field potential. Furthermore,
we include the in-medium modification of the width of the $\Delta$
resonance by taking into account that the decay might be Pauli
blocked and that there are additional channels for the $\Delta$ in
the medium like two and three body collisions which therefore yield
to a collisional broadening of resonances in the nuclear medium.
Including all these effects leads to a significant change of the
cross section compared to the vacuum result.

The third step is the propagation of the produced particles from
their production through the nuclear medium out to the detector.
During this propagation the particle originally produced can loose
parts of its energy and change its direction or even charge through
rescattering. It can also be absorbed, thus transferring its energy
and momentum to nucleons. These nucleons can then either be knocked
out of the nucleus or produce other, secondary hadrons in collisions
with other nucleons. The hadron ultimately seen leaving the nucleus
may thus not be that that was originally, in the first interaction
of the probe with a nucleon, produced.

The latter step is handled by a semiclassical coupled channel
transport theory with the help of the GiBUU code\cite{GiBUU} that
takes Fermi motion and Pauli blocking into account and allows for a
propagation of all hadrons in their mean field potentials.
Originally it has been developed for the description of heavy-ion
collisions and has since then been applied to - and tested against -
various more elementary reactions on nuclei with protons, pions,
elctrons, photons and neutrinos in the entrance channel. In this
method the spectral phase space distributions of all particles
involved are propagated in time, from the initial first contact of
the probe with the nucleus all the way to the final hadrons leaving
the nuclear volume on their way to the detector. The spectral phase
space distributions $F_h(\vec{r},\vec{p},\mu,t)$ give at each moment
of time and for each particle class $h$ the probability to find a
particle of that class with a (possibly off-shell) mass $\mu$ and
momentum $\vec{p}$ at position $\vec{r}$. Its time-development is
determined by the BUU equation
\begin{equation}     \label{BUU}
(\frac{\partial}{\partial t} + \frac{\partial H_h}{\partial \vec{p}}
\frac{\partial}{\partial \vec{r}} - \frac{\partial H_h}{\partial
\vec{r}} \frac{\partial}{\partial \vec{p}})F_h=G_h a_h - L_h F_h.
\end{equation}
Here $H_h$ gives the energy of the hadron $h$ that is being
transported; it contains the mass, the selfenergy (mean field) of
the particle and a term that drives an off-shell particle back to
its mass shell. The terms on the lhs of (\ref{BUU}) are the
so-called \emph{drift terms} since they describe the independent
transport of each hadron class $h$. The terms on the rhs of
(\ref{BUU}) are the \emph{collision terms}; they describe both
elastic and inelastic collisions between the hadrons. Here the term
\emph{inelastic collisions} includes those collisions that either
lead to particle production or particle absorption. The former is
described by the \emph{gain term} $G_h a_h$ on the rhs in
(\ref{BUU}), the latter process (absorption) by the \emph{loss term}
$L_h F_h$. Note that the gain term is proportional to the spectral
function $a$ of the particle being produced,
thus allowing for production of off-shell particles. On the
contrary, the loss term is proportional to the spectral phase space
distribution itself: the more particles there are the more can be
absorbed. The terms $G_h$ and $L_h$ on the rhs give the actual
strength of the gain and loss terms, respectively. They have the
form of Born-approximation collision integrals and take the
Pauli-principle into account. The free collision rates themselves
are taken from experiment or are calculated\cite{Effephot}.

The collision term on the rhs of (\ref{BUU}) is responsible for the
collision broadening that all particles experience when they are
embedded in a dense medium. Collisions either change energy and
momentum of the particles are absorb them alltogether. Both
processes contribute to collisional broadening. The detailed
structure of the gain and loss terms can be obtained from quantum
transport theory\cite{KadBaym,BotMal}.

\begin{figure}[tb]
\centerline{\includegraphics[width=13cm]{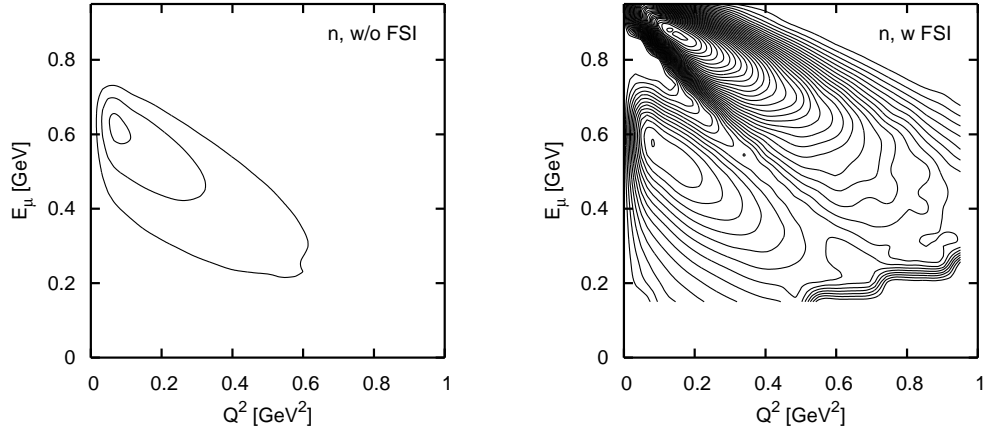}}
\caption{Double differential cross section $d\sigma/dQ^2\,dE_\mu$
for neutron knockout induced by charged current scattering of
$\nu_{\mu}$ on $^{56}$Fe at $E_\nu = 1$ GeV. Left: without FSI,
right: with FSI (from Ref.~\protect\refcite{Leitner-CC}).}
\label{n-knockout}
\end{figure}

A very dramatic example, which demonstrates the importance of
coupled channel effects, is provided by the charged current
neutrino-induced neutron knockout off nuclei. Since charged current
interactions by themselves always change the charge of the hit
nucleon by one unit there cannot be any charged current knock-out
neutrons in a quasielastic process. This is indeed born out in the
results of calculations (see Fig.~\ref{n-knockout},
left)\cite{Leitner-CC}. The few events visible in that picture at
$Q^2 \approx 0.05$ GeV$^2$ and $E_\mu \approx 0.6$ GeV stem from
events where first a $\Delta^+$ is produced that then decays into
$\pi^+ n$.

When final state interactions are turned on, this picture changes
dramatically (see Fig.~\ref{n-knockout}, right). Now a significant
neutron knockout signal appears at $E_\mu \approx 0.9$ GeV with a
long ridge in $Q^2$. In addition the $\Delta$-like events now show
also considerably more strength. The former effect is caused by
charge-transfer reactions where in a first interaction a proton is
knocked on that then travels through the nucleus and transmits its
energy and momentum to a hit neutron that is being knocked out of
the nucleus. The same applies to the $\Delta$-like events: due to
charge-exchange FSI now also the initial decay channels $\Delta^+
\to \pi^0 p$ and $\Delta^{++} \to \pi^+ p$ can contribute to final
neutrons being knocked out.

We shall now present two more applications, namely photon and
neutrino induced neutral current pion production on nuclei.

\section{Photoproduction of pions on nuclei}

\begin{figure}[tb]
\centerline{\includegraphics[width=15pc]{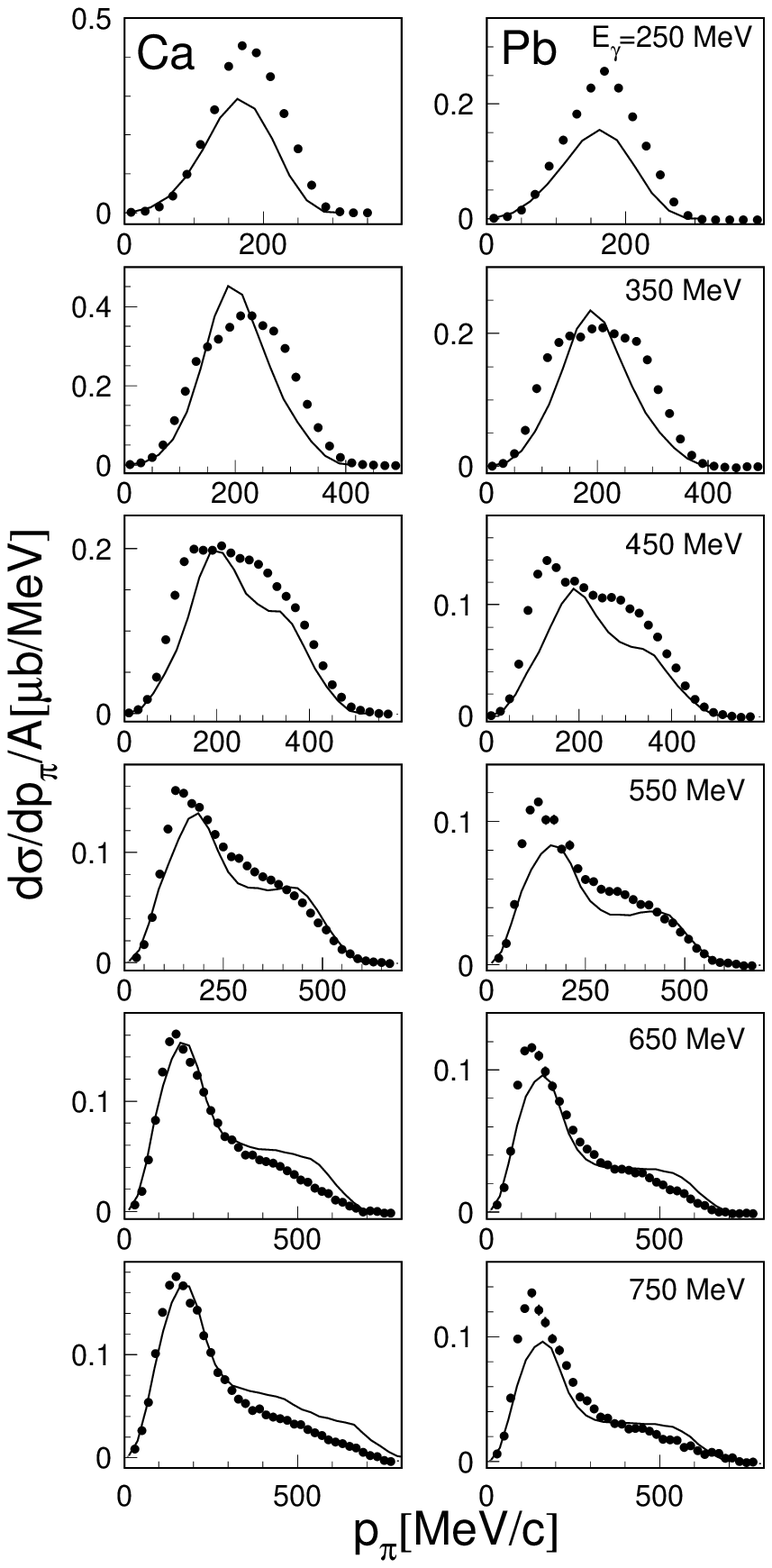}}
\caption{Photoproduction of $\pi^0$ on Ca (left) and Pb (right) as a
function of pion momentum. Shown is the BUU calculation in
comparison with data. The photon energies are given in the
individual frames (from Ref.~\protect\refcite{Krusche}).}
\label{pi0}
\end{figure}

An example for the method and the quality of its results is shown in
Fig.~\ref{pi0}.
Here we show the momentum-differential distributions for neutral
pions produced by real photons on the nuclei Ca and Pb. The overall
behavior of the spectra is described quite well by the BUU
calculations.  The clear deficiencies that show up at the lowest
photon energy of 250 MeV, where the calculated cross section is only
about 2/3 of the experimental one, is due to the fact that the data
here contain a significant contribution from coherent pion
production \cite{Krusche} which cannot be described by the transport
calculations. At the higher photon energies a distinct shape emerges
-- in agreement with experiment -- that reflects the $\pi N \Delta$
dynamics in nuclei. The spectra always start out at zero momentum
with zero cross section reflecting the $p$-state coupling of pions
to the $\Delta$. The following peak drops off steeply at momenta of
around 200 MeV reflecting the strong pion absorption through the
$\Delta$ resonance. After the fall-off the spectrum flattens and
smoothly decreases to zero, as mandated by phase-space limitations.
The structure just described shows up in the data and the
calculations as well only for photon energies above about 450 MeV
where the $\Delta$ resonance is well excited.

\section{Neutrino induced neutral current pion production}

\begin{figure}[tb]
\centerline{\includegraphics[width=13cm]{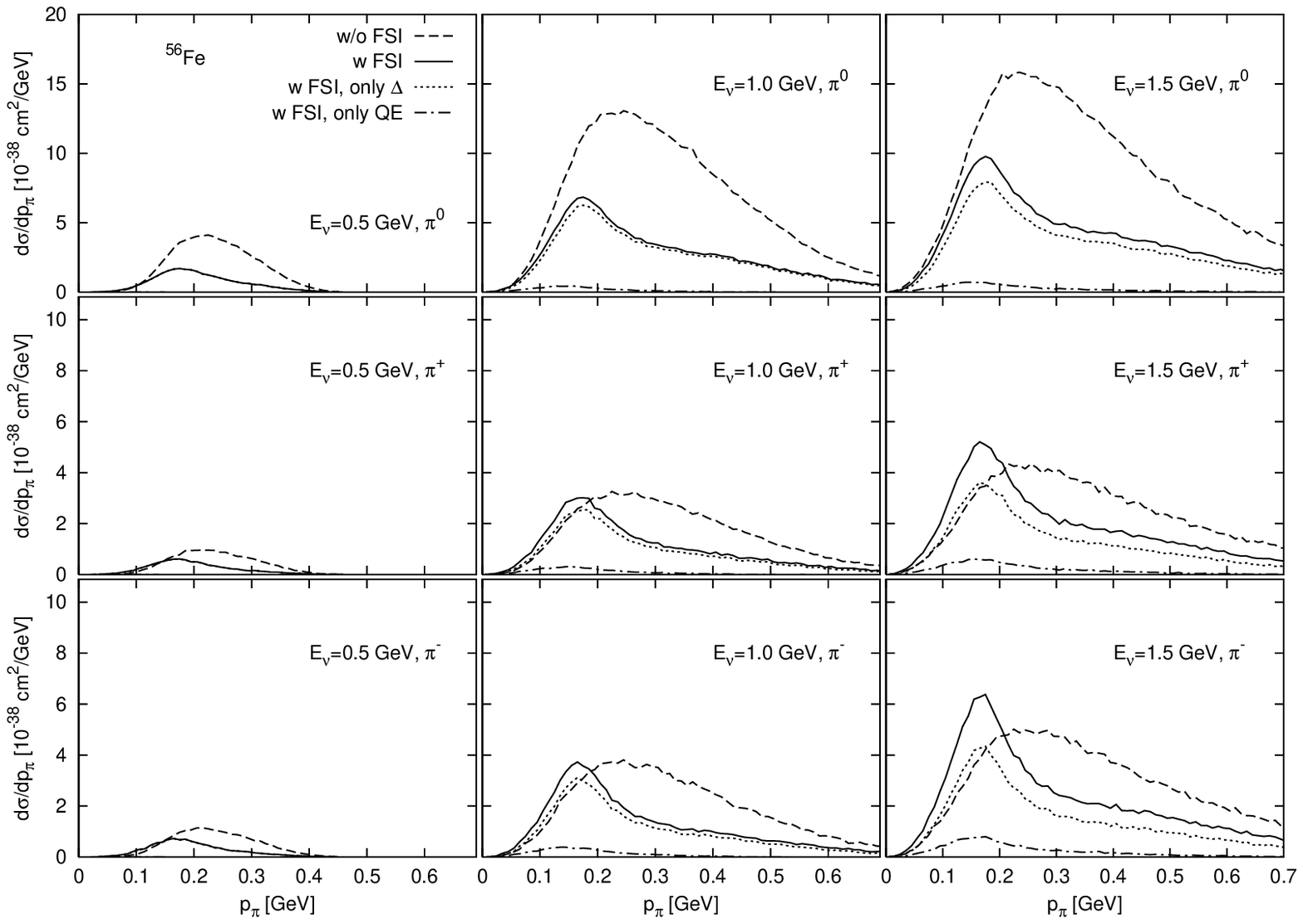}}
\caption{Momentum differential cross section for $\pi$ production on
$^{56}\text{Fe}$ versus the pion momentum $p_{\pi}$ at different
values of $E_{\nu}$. The dashed lines denote the calculation without
FSI; the solid lines denote the one with FSI. Furthermore, it is
indicated whether the pion comes from initial QE or $\Delta$
excitation (dash-dotted or dotted lines).
(cf.~Ref.~\protect\refcite{Leitner-NC}).} \label{NC_mom_spectra}
\end{figure}

Exactly the same behavior as for photoproduction of pions on nuclei
also shows up in the neutrino-induced pion production from nuclear
targets\cite{Leitner-CC,Leitner-NC}. We illustrate this in
Fig.~\ref{NC_mom_spectra} with the momentum-differential spectrum of
pions produced by neutral current scattering of neutrinos on
$^{56}$Fe for 3 neutrino energies.
While the overall shape of the result without FSI (dashed line) is
again dictated by the predominant $p$-wave production mechanism
through the $\Delta$ resonance, the shape of the solid lines which
denote the full calculation is influenced by the energy dependence
of the pion absorption and rescattering. The main absorption
mechanism for pions above $p_{\pi}\approx 0.2$~GeV is $\pi N \to
\Delta$ followed by $\Delta N \to N N$ which leads to a considerable
reduction of the cross section. Elastic scattering $\pi N \to \pi N$
redistributes the kinetic energies and thus also shifts the spectrum
to lower energies.

While this is equivalent to the photoproduction case, we want to
point out an interesting feature specific to neutrino reactions. As
a direct consequence of the isospin structure of the resonance
decay, the cross section for $\pi^0$~production is significantly
higher than those of the $\pi^+$ and $\pi^-$ channels. When FSI are
included, we find an enhancement of the peaks in the middle and
bottom panels of Fig.~\ref{NC_mom_spectra} over the value obtained
without FSI. This is due to the fact that the $\pi^0$ undergo charge
exchange and contribute to the charged channels (side-feeding). The
effect in the opposite direction is less important due to the
smaller elementary $\pi^+$ and $\pi^-$ production cross section.

Pions can also emerge from the initial QE neutrino-nucleon reaction
when the produced nucleon rescatters producing a $\Delta$ or
directly a pion (see dash-dotted line). This contributes mostly to
the low energy region of the pion spectra due to the redistribution
of the energy in the collisions. However, this process is not very
sizable because it is relevant only at high $Q^2$.

\begin{figure}[tb]
\centerline{\includegraphics[width=7cm]{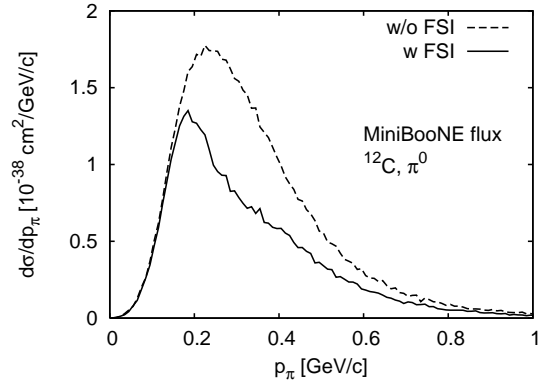}}
\caption{Momentum differential cross section for NC $\pi^0$
production on $^{12}$C calculated with the MiniBooNE incoming
neutrino energy distribution.} \label{miniboone}
\end{figure}

Finally, we show in Fig.~\ref{miniboone} our result for the neutral
current $\pi^0$ production on $^{12}$C. Plotted is the momentum
differential cross section versus the pion momentum averaged over
the incoming neutrino energy distribution of the MiniBooNE
experiment\cite{miniboone} as given in Ref.~\refcite{Monroe}. In
principle, our model allows for the inclusion of detector
acceptances, however, it is not considered in this calculation.
The dashed line shows the spectrum including Fermi motion and Pauli
blocking, but no FSI, the solid curve gives the spectrum with the
FSI turned on.
Again, we find, that the shape of the spectrum changes significantly.

\section{Summary}

In this talk various aspects of in-medium effects have been
demonstrated. Any in-medium signal that involves hadrons in the
final states is subject to final state interactions, thus, for a
reliable predictions of observables one has to take these final
state interactions with all their complications in a coupled channel
calculation into account; simple Glauber-type descriptions are not
sufficient. It was outlined that transport theory is at present the
only reliable method to calculate the observable consequences of
in-medium properties of hadrons and their interactions; usable
quantum-mechanical approaches for the description of semi-inclusive
events do not exist. Special emphasis was put on the demonstration
of the overwhelming influence of final state interactions using
examples from photon-nucleus and neutrino-nucleus interactions.

\section{Acknowledgements}
This work has been supported by DFG and BMBF.

\end{document}